\documentclass[12pt]{article}

\usepackage{amsfonts}
\usepackage{amsmath}
\usepackage{amssymb}
\usepackage{graphicx}
\usepackage{pst-all}

\newtheorem{theorem}{\bf Theorem}[section]
\newtheorem{corollary}[theorem]{\bf Corollary}
\newtheorem{lemma}[theorem]{\bf Lemma}

\newcommand{\proof}{\noindent{\bf Proof.\ }}
\newcommand{\qed}{\hfill $\square$ \bigskip}

\begin{document}

\title{Cycles in enhanced hypercubes}

\author{
Meijie Ma \\
Department of Mathematics  \\
Zhejiang Normal University   \\
Jinhua, Zhejiang, 321004, China \\
{mameij@mail.ustc.edu.cn} }
\date{}
\maketitle

\begin{abstract}
The enhanced hypercube $Q_{n,k}$ is a variant of the hypercube
$Q_n$. We investigate all the lengths of cycles that an edge of the
enhanced hypercube lies on. It is proved that every edge of
$Q_{n,k}$ lies on a cycle of every even length from $4$ to $2^n$; if
$k$ is even, every edge of $Q_{n,k}$ also lies on a cycle of every
odd length from $k+3$ to $2^n-1$, and some special edges lie on a
shortest odd cycle of length $k+1$.
\end{abstract}



\baselineskip18pt

\section{Introduction}
\label{sec:intro}
As a topology for an interconnection network of a multiprocessor
system, the hypercube structure is a widely used and well-known
interconnection model since it possesses many attractive
properties~\cite{ss88, x01}. In particular, $Q_n$ is
vertex-transitive and edge-transitive. There are different
variations of hypercubes, for example, folded
hypercubes~\cite{as91}, balanced hypercubes~\cite{e91}, and
augmented hypercubes~\cite{cs2002}. These variations support
efficient embedding, reduced diameter, and improve fault tolerance
in comparison to the hypercube.

The enhanced hypercube $Q_{n,k}$ was introduced by Tzeng and
Wei~\cite{tw1991} to achieve improvement in diameter, mean internode
distance, and traffic density. Different properties of enhanced
hypercubes were investigated by now.  The diagnosability of it under
different models was studied in~\cite{w1994, w1999}. To measure the
fault-tolerance of them, the super spanning connectivity and super
spanning laceability were determined in~\cite{clthh2009}. The
authors constructed independent spanning trees in~\cite{ycpc2014}.
Paths and cycles embedding with faulty elements were investigated
in~\cite{ll2012, ll2013, llq2013, ql2014}. These results indicate
that enhanced hypercubes keep numerous desirable properties of
hypercubes.

A cycle structure, which is a fundamental topology for parallel
and distributed processing, is suitable for local area networks
and for the development of simple parallel algorithms with low
communication cost.  Studies of various networks about the cycle
embedding problems can be found in the literature~\cite{ch2014,
ch2015, ml2009, wllw2013}. To the best of our knowledge, there
are no results about cycle embedding with respect to a specific
edge in $Q_{n,k}$. The enhanced hypercube $Q_{n,k}$ is vertex
transitive~\cite{ycpc2014}, but not edge
transitive~\cite{tw1991} except for $Q_{n,n}$ which is the
folded hypercube $FQ_n$~\cite{xmd06}. As can be seen from the
proofs, the embedding even cycles containing a special edge is
straightforward when compared with the more interesting case of
embedding the shortest odd cycles.

The rest of this paper is organized as follows. In the next
section we introduce the basic notation and terminology. The
main results are given in Section~\ref{sec:property}.

\section{Preliminaries}
\label{sec:exchanged}

Throughout this paper, we follow Xu~\cite{x01} for graph-theoretical
terminology and notation not defined here.

A $uv$-walk $W$ in a graph $G$ is a sequence of vertices in $G$,
beginning with $u$ and ending at $v$ such that consecutive vertices
in the sequence are adjacent. That is, we can express $W$ as
$W=\langle u=v_0,v_1,\ldots, v_l=v \rangle$ where $l\geq 0$ and
$(v_i,v_{i+1})\in E(G)$ for $0\leq i\leq l-1$. Each vertex $v_i$
$(0\leq i\leq l)$ and each edge $(v_i,v_{i+1})$ $(0\leq i\leq l-1)$
is said to lie on $W$. The number of edges in a walk is called the
length of the walk. If $u=v$, then the walk is closed; if $u\neq v$,
then $W$ is open.  A $uv$-path $P$ in a graph $G$ is a $uv$-walk in
which no vertices are repeated. A closed walk that repeats no
vertices, except for the first and last, is a cycle. The length of
any cycle is at least $3$. We usually express a cycle as
$C=P_1+\langle v_0,v_1,\ldots, v_i \rangle$ where $P_1$ is a
$v_0v_i$-path disjoint with path $\langle v_0,v_1,\ldots, v_i
\rangle$ except for the two vertices $v_0$ and $v_i$.

The $n$-dimensional hypercube $Q_n$ is a graph with $2^n$
vertices, each vertex with a distinct binary string
$u_nu_{n-1}\cdots u_1$ on the set $\{0,1\}$. Two vertices are
linked by an edge if and only if their strings differ in exactly
one bit. The edge joining the vertices differ in the $i$-th
($1\leq i\leq n$) bit is called an $i$-dimensional edge. The
Cartesian product $G \times H$ of graphs $G$ and $H$ is the
graph with vertex set $V(G)\times V(H)$, vertices $(g,h)$ and
$(g',h')$ being adjacent whenever $gg'\in E(G)$ and $h=h'$, or
$g=g'$ and $hh'\in E(H)$. For our purposes it is essential to
recall that $Q_n$ can be represented as $Q_n=K_2 \times
Q_{n-1}$.


\begin{figure}[h!]
\label{fig:EQ1}
\begin{pspicture}(-0.6,-.5)(3,4)
\psset{radius=.08, unit=.8}

\Cnode(1,1){000}\rput(.6,0.75){\tiny000}
\Cnode(3.5,1){010}\rput(3.8,.75){\tiny010}
\Cnode(1,3.5){100}\rput(.5,3.5){\tiny100}
\Cnode(3.5,3.5){110}\rput(3.4,3.8){\tiny110}

\Cnode(2.3,1.8){001}\rput(2.5,1.6){\tiny001}
\Cnode(4.8,1.8){011}\rput(5.25,1.98){\tiny011}
\Cnode(2.3,4.3){101}\rput(2.3,4.6){\tiny101}
\Cnode(4.8,4.3){111}\rput(4.8,4.6){\tiny111}

\ncline[linewidth=1.pt]{000}{100} \ncline[linewidth=1.pt]{001}{011}
\ncline[linewidth=1.pt]{101}{111} \ncline[linewidth=1.pt]{011}{111}

\ncline[linewidth=1.pt]{001}{101} \ncline[linewidth=1.pt]{000}{010}
\ncline[linewidth=1.pt]{100}{110} \ncline[linewidth=1.pt]{010}{110}

\ncline[linewidth=1.pt]{010}{011} \ncline[linewidth=1.pt]{110}{111}
\ncline[linewidth=1.pt]{100}{101} \ncline[linewidth=1.pt]{000}{001}

\ncline[linestyle=dashed, dash=3pt
2pt]{101}{010}\ncline[linestyle=dashed, dash=3pt 2pt]{100}{011}

\ncline[linestyle=dashed, dash=3pt
2pt]{000}{111}\ncline[linestyle=dashed, dash=3pt 2pt]{001}{110}

\rput(2.3,0.1){\scriptsize (a) $Q_{3,3}$}
\end{pspicture}
\begin{pspicture}(-1.6,-.5)(6,4)
\psset{radius=.08, unit=.8}

\Cnode(1,1){0000}\rput(.6,0.75){\tiny0000}
\Cnode(3.5,1){0010}\rput(3.8,.75){\tiny0010}
\Cnode(1,3.5){0100}\rput(.5,3.5){\tiny0100}
\Cnode(3.5,3.5){0110}\rput(3.36,3.75){\tiny0110}

\Cnode(2.3,1.8){0001}\rput(2.5,1.6){\tiny0001}
\Cnode(4.8,1.8){0011}\rput(5.25,1.98){\tiny0011}
\Cnode(2.3,4.3){0101}\rput(2.3,4.6){\tiny0101}
\Cnode(4.8,4.3){0111}\rput(4.8,4.6){\tiny0111}

\ncline[linewidth=1.pt]{0000}{0100}
\ncline[linewidth=1.pt]{0001}{0011}
\ncline[linewidth=1.pt]{0101}{0111}
\ncline[linewidth=1.pt]{0011}{0111}

\ncline[linewidth=1.pt]{0001}{0101}
\ncline[linewidth=1.pt]{0000}{0010}
\ncline[linewidth=1.pt]{0100}{0110}
\ncline[linewidth=1.pt]{0010}{0110}

\ncline[linewidth=1.pt]{0010}{0011}
\ncline[linewidth=1.pt]{0110}{0111}
\ncline[linewidth=1.pt]{0100}{0101}
\ncline[linewidth=1.pt]{0000}{0001}

\ncline[linestyle=dashed, dash=3pt
2pt]{0101}{0010}\ncline[linestyle=dashed, dash=3pt 2pt]{0100}{0011}

\ncline[linestyle=dashed, dash=3pt
2pt]{0000}{0111}\ncline[linestyle=dashed, dash=3pt 2pt]{0001}{0110}

\Cnode(6,1){1010}\rput(5.6,0.75){\tiny1010}
\Cnode(8.5,1){1000}\rput(8.8,.75){\tiny1000}
\Cnode(6,3.5){1110}\rput(6.7,3.65){\tiny1110}
\Cnode(8.5,3.5){1100}\rput(8.4,3.8){\tiny1100}

\Cnode(7.3,1.8){1011}\rput(7.5,1.6){\tiny1011}
\Cnode(9.8,1.8){1001}\rput(10.25,1.8){\tiny1001}
\Cnode(7.3,4.3){1111}\rput(7.3,4.6){\tiny1111}
\Cnode(9.8,4.3){1101}\rput(9.8,4.6){\tiny1101}

\ncline[linewidth=1.pt]{1000}{1100}
\ncline[linewidth=1.pt]{1001}{1011}
\ncline[linewidth=1.pt]{1101}{1111}
\ncline[linewidth=1.pt]{1011}{1111}

\ncline[linewidth=1.pt]{1001}{1101}
\ncline[linewidth=1.pt]{1000}{1010}
\ncline[linewidth=1.pt]{1100}{1110}
\ncline[linewidth=1.pt]{1010}{1110}

\ncline[linewidth=1.pt]{1010}{1011}
\ncline[linewidth=1.pt]{1110}{1111}
\ncline[linewidth=1.pt]{1100}{1101}
\ncline[linewidth=1.pt]{1000}{1001}

\ncline[linewidth=1.pt]{1010}{0010}
\ncline[linewidth=1.pt]{1011}{0011}
\ncline[linewidth=1.pt]{1110}{0110}
\ncline[linewidth=1.pt]{1111}{0111}

\ncline[linestyle=dashed, dash=3pt
2pt]{1101}{1010}\ncline[linestyle=dashed, dash=3pt 2pt]{1100}{1011}

\ncline[linestyle=dashed, dash=3pt
2pt]{1000}{1111}\ncline[linestyle=dashed, dash=3pt 2pt]{1001}{1110}

\nccurve[angleA=17,angleB=163]{0101}{1101}
\nccurve[angleA=17,angleB=163]{0100}{1100}
\nccurve[angleA=17,angleB=163]{0001}{1001}
\nccurve[angleA=17,angleB=163]{0000}{1000}

\rput(4.8,0.1){\scriptsize (b) $Q_{4,3}$}

\end{pspicture}
\caption{Enhanced hypercubes $Q_{3,3}$ and $Q_{4,3}$. The hypercube
edges and skips are represented by solid lines and dashed lines,
respectively.}
\end{figure}
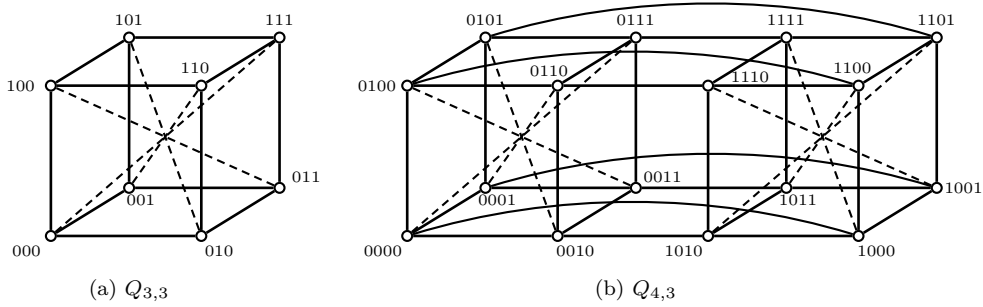

As a variant of the hypercube, the $n$-dimensional folded hypercube
$FQ_n$, proposed first by El-Amawy and Latifi~\cite{as91}, can be
obtained from the hypercube $Q_n$ by adding an edge, called a {\it
complementary edge}, between any pair of complementary vertices
$u=u_n\ldots u_2u_1$ and
$\bar{u}=\bar{u}_n\ldots\bar{u}_2\bar{u}_1$, where $\bar{u}_i=1-
u_i$ for $i=1,2,\ldots,n$.

The $n$-dimensional enhanced hypercube $Q_{n,k}$ ($1\leq k\leq
n$) is obtained from the hypercube $Q_n$ by adding a
complementary edge between two vertices $u=u_n\ldots u_2u_1$ and
$v=u_n\ldots u_{k+1}\bar{u}_k\bar{u}_{k-1}\ldots \bar{u}_1$. We
call these complementary edges skips, and denote by $E_s$. To
distinguish $E_s$ from the edges in $Q_n$, we call edges in
$Q_n$ hypercube edges and denote the set of $i$-dimensional
hypercube edges by $E_i$ for $i=1,2,\ldots,n$.  So the complete
edge set $E(Q_{n,k})$ of a enhanced hypercube can be expressed
as $E(Q_n)\cup E_s$. For instance, $Q_{3,3}$ and $Q_{4,3}$ are
shown in Fig.~1.

When $k=1$,  $Q_{n,1}$ reduces to the $n$-dimensional hypercube
$Q_n$. When $k=n$,  $Q_{n,n}$ is the $n$-dimensional folded
hypercube $FQ_n$. Since there are many results about cycle
embedding in the hypercubes $Q_n$ and folded hypercubes
$FQ_n$~\cite{chf2013,chf2015, lt03, xm06, xmd06}. We consider
cycles in the enhanced hypercubes $Q_{n,k}$ where $2\leq k\leq
n-1$ in the next section.

From the definitions of Cartesian product graph and the enhanced
hypercube, we have $Q_{n,k}=K_2\times Q_{n-1,k}$ when $1\leq
k\leq n-1$. Thus $Q_{k,k}$ is the folded hypercube $FQ_k$, and
$Q_{n,k}$ $(n \geq k+1)$ is built from two copies of $Q_{n-1,k}$
as follows: Let $x \in \{0, 1\}$ and let $xQ_{n-1,k}$ denote the
graph obtained by prefixing the label of each node of one copy
of $Q_{n-1,k}$ with $x$; connect each node $0u_{n-1}\ldots
u_2u_1$ of $0Q_{n-1,k}$ with the node $1u_{n-1}\ldots u_2u_1$ of
$1Q_{n-1,k}$ by an edge.

\section{Cycles embedding in $Q_{n,k}$}
\label{sec:property}

For convenience, we use $0G$ and $1G$ to denote the two graphs
isomorphism to $G$, respectively. Then, $K_2\times G$ can be
obtained from $0G$ and $1G$ by joining $0u$ and $1u$ between $0G$
and $1G$ for any $u\in V(G)$. The edges between $0G$ and $1G$ are
called crossing edges.

We first prove some lemmas about the graph $K_2\times G$.

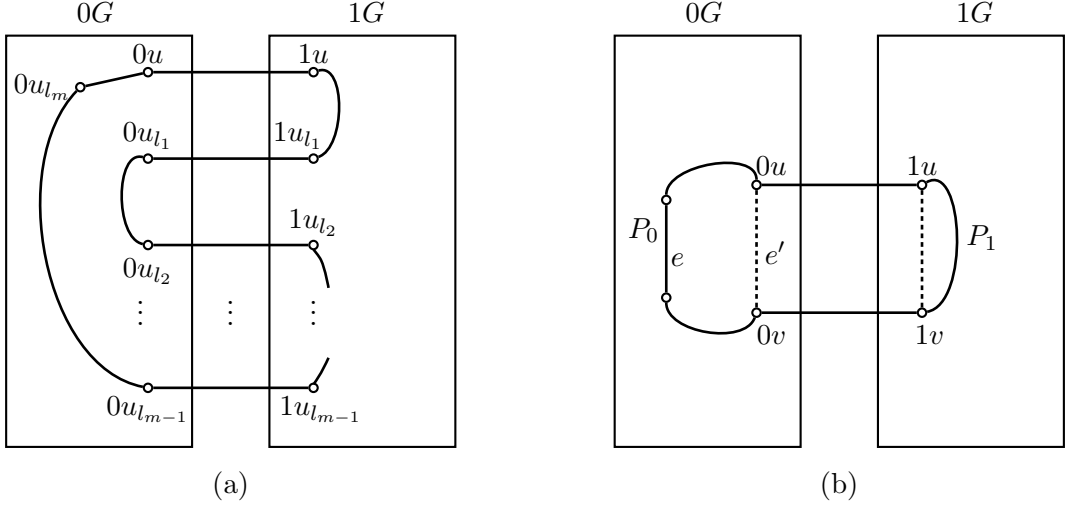
\begin{figure}
\begin{pspicture}(0,0)(8,6.5)
\psset{radius=.07}

\psframe(0.5,1)(3,6.5) \psframe(4,1)(6.5,6.5)
\Cnode(2.4,6){1u}\rput(2.4,6.25){\small$0u$}
\Cnode(4.6,6){0u}\rput(4.6,6.25){\small$1u$}
\Cnode(2.4,4.85){1u2}\rput(2.4,5.15){\small$0u_{l_1}$}
\Cnode(4.6,4.85){0u2}\rput(4.4,5.15){\small$1u_{l_1}$}

\Cnode(2.4,3.7){1u3}\rput(2.4,3.35){\small$0u_{l_2}$}
\Cnode(4.6,3.7){0u3}\rput(4.6,4){\small$1u_{l_2}$}
\Cnode(4.6,1.8){0u4}\rput(4.7,1.5){\small$1u_{l_{m-1}}$}
\Cnode(2.4,1.8){1u4}\rput(2.4,1.5){\small$0u_{l_{m-1}}$}
\Cnode(1.5,5.8){1u5}\rput(1.,5.8){\small$0u_{l_{m}}$}

\psset{linewidth=1.pt}
\ncline{1u}{0u}\ncline{1u2}{0u2}\ncline{1u3}{0u3} \ncline{0u4}{1u4}
\ncline{1u5}{1u} \nccurve[angleA=17,angleB=19]{0u}{0u2}
\nccurve[angleA=164,angleB=169]{1u2}{1u3}
\nccurve[angleA=-134,angleB=169]{1u5}{1u4}

\rput(2.3,2.9){$\vdots$} \rput(4.6,2.9){$\vdots$}
\rput(3.5,2.9){$\vdots$}

\pscurve(4.6,3.64)(4.7,3.5)(4.8,3.13)
\pscurve(4.6,1.845)(4.7,2.0)(4.8,2.2)

\rput(1.7,6.8){\small$0G$}\rput(5.3,6.8){\small$1G$}
\rput(3.5,0.5){\small (a)}
\end{pspicture}
\begin{pspicture}(0,0)(8,6.5)
\psset{radius=.07}

\psframe(0.5,1)(3,6.5) \psframe(4,1)(6.5,6.5)
\Cnode(2.4,4.5){1u}\rput(2.6,4.75){\small$0u$}
\Cnode(4.6,4.5){0u}\rput(4.6,4.75){\small$1u$}

\Cnode(4.6,2.8){0u4}\rput(4.7,2.5){\small$1v$}
\Cnode(2.4,2.8){1u4}\rput(2.6,2.5){\small$0v$}

\Cnode(1.2,4.3){1} \Cnode(1.2,3){2}\rput(.9,3.9){\small$P_0$}
\rput(1.35,3.5){\small$e$} \rput(2.65,3.6){\small$e'$}

\psset{linewidth=1.pt} \ncline{1}{2}
\ncline{1u}{0u}\ncline{0u4}{1u4}\ncline[linestyle=dashed,dash=1.8pt
1.5pt]{1u}{1u4} \ncline[linestyle=dashed,dash=1.8pt 1.5pt]{0u}{0u4}
\nccurve[angleA=30,angleB=10]{0u}{0u4}\nccurve[angleA=90,angleB=90]{1}{1u}
\nccurve[angleA=270,angleB=250]{2}{1u4}

\rput(5.4,3.8){\small$P_1$}
\rput(1.7,6.8){\small$0G$}\rput(5.3,6.8){\small$1G$}

\rput(3.5,0.5){\small (b)}
\end{pspicture}

\caption{\label{f2}\footnotesize{Illustrations for the proof of
Lemma \ref{l1} and \ref{ll}.(A straight or dashed line represents an
edge, a curved line represents a path between two vertices.) }}
\end{figure}

\begin{lemma}
\label{l1} If the length of a shortest odd cycle in a graph $G$
is $l$, the length of a shortest odd cycle containing a crossing
edge is at least $l+2$ in $K_2\times G$.
\end{lemma}

\proof Assume the crossing edge  is $(0u,1u)$ in $K_2\times G$.

We will prove the length of any odd cycle containing the edge
$(0u,1u)$ is at least $l+2$. Assume the odd cycle is $C=\langle
0u, 1u,
1u_{1},\ldots,1u_{l_1},0u_{l_1},0u_{l_1+1},\ldots,0u_{l_2},1u_{l_2},\ldots,
\\1u_{l_3},0u_{l_3},\ldots,0u_{l_m},0u\rangle$ (See Fig.\ref{f2}(a)).
Then the length of the cycle $C$ is $l_m+1+m$ and $m\geq 2$ is
an even integer. A closed walk $W$ in $G$ is obtained by
removing the left bits of the vertices on $C$. The closed walk
$W$ is $\langle u,
u_{1},\ldots,u_{l_1},\ldots,u_{l_2},\ldots,u_{l_3},\ldots,u_{l_m},u
\rangle$. The length of the closed walk $W$ is  $l_m+1$. Since
$l_m+1+m$ is odd and $m$ is even, $l_m+1$ is odd. Since any odd
closed walk must contain an odd cycle, and the length of any odd
cycle in $G$ is at least $l$, we have $l_m+1\geq l$. Hence
$l_m+1+m\geq l+2$, that is the length of the odd cycle
containing the edge $(0u,1u)$ is at least $l+2$. \qed

From Lemma~\ref{l1}, the following conclusion is obvious.

\begin{corollary}\label{c1}
If the length of a shortest odd cycle in graph $G$ is $l$, then
the length of a shortest odd cycle in $K_2\times G$  is also
$l$.
\end{corollary}

\begin{lemma}
\label{ll} If every edge of a connected graph $G$ lies on a
cycle of length $l$ for any $l\in I$ ($I$ is a set of integers),
then every edge of $K_2\times G$ lies on a cycle of length $l'$
where $l'\in \{l+2|l\in I\}\cup \{l_1+l_2|l_1, l_2\in I\}$.
\end{lemma}

\proof The graph $K_2\times G$ can be obtained from $0G$ and $1G$ by
adding crossing edges. Consider the following two cases.

{\it Case 1}\quad The edge $e$ lies in a subgraph $0G$ or $1G$.
Assume $e$ lies in $0G$.

If $l'=l+2$, there is a cycle  $C_0$ of length $l$ containing
$e$ in $0G$. Since $3\leq l$, there is an edge $e'=(0u,0v)$
different with $e$ on the cycle $C_0$. Let $P_0$ be the path
obtained by deleting the edge $e'$  from $C_0$. Then
$C=P_0+\langle 0u,1u,1v,0v\rangle$ is a cycle of length $l'+2$
containing $e$ in $K_2\times G$.

If $l'=l_1+l_2$, there is a cycle  $C_0$ of length $l_1$
containing $e$  in $0G$. Since $3\leq l$, there is an edge
$e'=(0u,0v)$ different with $e$ on the cycle $C_0$. Let $P_0$ be
the path obtained by deleting the edge $e'$  from $C_0$. There
is a cycle $C_1$ of length $l_2$ in $1G$ containing the edge
$(1u,1v)$. Let $P_1$ be the path obtained by deleting the edge
$(1u,1v)$  from $C_1$. Then $C=P_0+(0u,1u)+P_1+(1v,0v)$ is a
cycle of length $l$ containing $e$ (See Fig.\ref{f2}(b)).

{\it Case 2}\quad The edge $e=(0u,1u)$ is a crossing edge of
$K_2\times G$.

There is a neighbor $0v$ of $0u$ in $0G$, and $1v$ and $1u$ are also
adjacent in $1G$. Then $C=\langle 0u, 1u, 1v, 0v, 0u\rangle$ is a
cycle of length $4$ containing $e$.

If $l'=l+2$, there is a cycle  $C_0$ of length $l$ containing the
edge $(0u,0v)$ in $0G$.  Let $P_0$ be the path obtained by deleting
the edge $(0u,0v)$  from $C_0$. Then $C=P_0+\langle
0u,1u,1v,0v\rangle$ is a cycle of length $l'+2$ containing
$e=(0u,1u)$ in $K_2\times G$.

If $l'=l_1+l_2$, there is a cycle  $C_0$ of length $l_1$ containing
the edge $(0u,0v)$ in $0G$.  Let $P_0$ be the path obtained by
deleting the edge $(0u,0v)$  from $C_0$. There is a cycle $C_1$ of
length $l_2$ in $1G$ containing the edge $(1u,1v)$. Let $P_1$ be the
path obtained by deleting the edge $(1u,1v)$  from $C_1$. Then
$C=P_0+(0u,1u)+P_1+(1v,0v)$ is a cycle of length $l$ containing
$e=(0u,1u)$.

The lemma follows. \qed

The following two lemmas which obtained in the literature make the
proofs of our results simple.

\begin{lemma} (\cite{lt03})\label{l2}
Every edge of $Q_n$ lies on a cycle of every even length from $4$ to
$2^n$ for $n\ge 2$.
\end{lemma}

\begin{lemma} (\cite{xm06})\label{l3}
The $n$-dimensional folded hypercube $FQ_n$ is bipartite if and only
if $n$ is odd, and the minimum length of odd cycles is $n+1$ if $n$
is even. Every edge of $FQ_n$ lies on a cycle of every even length
from $4$ to $2^n$ for $n\ge 3$. Moreover, every edge also lies on a
cycle of every odd length from $n+1$ to $2^n-1$ if $n$ is even.
\end{lemma}

\begin{lemma}
The enhanced hypercube $Q_{n,k}$ is a bipartite graph if and only if
$k$ is odd.
\end{lemma}

\proof Since $Q_{n,k}$ is obtained from $Q_n$ by adding
$2^{n-1}$ skips, to prove the lemma, it is sufficient to
consider skips. Let $\{X,Y\}$ be a bipartition of $Q_n$. Since
any vertex $u=u_n\ldots u_2u_1$ and $v=u_n\ldots
u_{k+1}\bar{u}_k\bar{u}_{k-1}\ldots \bar{u}_2\bar{u}_1$ in $Q_n$
have different parity if and only if $k$ is odd and, hence, any
skip in $Q_{n,k}$ joins two vertices in different parts of
$\{X,Y\}$. It follows that $Q_{n,k}$ is a bipartite graph if and
only if $k$ is odd. \qed

\begin{theorem}
\label{t1} Every edge of $Q_{n,k}$ ($1\leq k\leq n$) lies on a
cycle of every even length from $4$ to $2^n$ for $n\ge 3$.
Moreover, every edge lies on a cycle of every odd length from
$k+3$ to $2^n-1$ if $k$ is even, and the edge not in  $E_i$
($k+1\leq i\leq n$) also lies on an odd cycle of length $k+1$.
\end{theorem}

\proof By Lemmas~\ref{l2} and~\ref{l3}, the conclusion is true for
$Q_{n,1}=Q_n$ and $Q_{n,n}=FQ_n$. Assume $2\leq k\leq n-1$. We prove
the theorem by induction on $n-k$.

When $n-k=1$  and  $n\geq 3$, that is $k=n-1\geq 2$. We have
$Q_{n,n-1}=K_2\times FQ_n$. Note that $FQ_2$ is a complete graph
$K_4$. By Lemmas~\ref{ll} and~\ref{l3}, the conclusion is true.

Assume that the conclusion is true for a integer $t$ with $1\leq
n-k< t$. We will prove the case $n-k=t$. Note that
$Q_{n,k}=K_2\times Q_{n-1,k}$.

By the induction hypothesis and Lemma~\ref{ll}, every edge of
$Q_{n,k}$ lies on a cycle of even length $l$ where $4\leq l\leq
2^n$.

Assume $k$ is even in the following. By Corollary~\ref{c1} and
Lemma~\ref{l3}, the length of the shortest odd cycle in $Q_{n,k}$ is
$k+1$.

If the edge $e=(0u,1u)$ is in $E_n$, that is $e$ is a crossing edge.
By the induction hypothesis and Lemma~\ref{l1}, the length of the
shortest odd cycle containing a crossing edge $e$ is at least $k+3$.
Since $k+3=2+(k+1)$, and there is a $1$-dimensional edge $(0u,0v)$
incident with $0u$ in $0Q_{n-1,k}$. By the induction hypothesis,
there is a cycle $C_0$ of length $k+1$ containing the edge $(0u,0v)$
in $0Q_{n-1,k}$. Let $P_0$ be the path obtained by deleting the edge
$(0u,0v)$  from $C_0$. Then $C=P_0+\langle 0u,1u,1v,0v\rangle$ is a
cycle of length $k+3$ containing $e=(0u,1u)$ in $Q_{n,k}$. For any
odd integer $k+5\leq l\leq 2^n-1$, we can express $l=l_1+l_2$ where
$k+1\leq l_1\leq 2^{n-1}-1$ is odd, and $4\leq l_2\leq 2^{n-1}$ is
even. By Lemma~\ref{ll}, there is an odd cycle of length $l$
containing $e$ in $Q_{n,k}$.

If the edge $e$ is not in $E_n$, that is $e$ is not a crossing
edge. We may assume that $e$ is in $0Q_{n-1,k}$. By the
induction hypothesis, the edge $e$ lies on an odd cycle of
length $l$ ($k+3\leq l\leq 2^{n-1}-1$), and the edge $e$ also
lies on an odd cycle of length $k+1$ if  $e\notin E_i$ ($k+1\leq
i\leq n-1$). For any odd integer $2^{n-1}+1 \leq l\leq 2^n-1$,
we can express $l=l_1+l_2$ where $k+3\leq l_1\leq 2^{n-1}-1$ is
odd, and $4\leq l_2\leq 2^{n-1}$ is even. By Lemma~\ref{ll},
there is an odd cycle of length $l$ containing $e$ in $Q_{n,k}$.
\qed

By Lemma~\ref{l1}, Corollary~\ref{c1}, and Theorem~\ref{t1}, we
conclude that the length of a shortest odd cycle is $k+1$, and the
length of a shortest odd cycle containing the edge of dimension $i$
($k+1\leq i\leq n$) is $k+3$ in $Q_{n, k}$ when $k$ is even. Hence,
the results are best possible.

Since every vertex of $Q_{n,k}$ is incident with a unique edge
in $E_i$ where $1\leq i\leq n$, we obtain the following
corollary.

\begin{corollary}
\label{cor:vertex} Every vertex of $Q_{n,k}$ lies on a cycle of
every even length from $4$ to $2^n$; if $k$ is even, every vertex
also lies on a cycle of every odd length from $k+1$ to $2^n-1$.
\end{corollary}

\section*{Acknowledgements}

This work was supported by the national natural science
foundation of China grant 11101378 and 11571319, and ZJNSF (No.
LY14A010009).


\end{document}